\documentclass[12pt]{article}
\usepackage{a4}
\usepackage{epsf}
\usepackage{amssymb}
\usepackage{cite}
\newcommand{\be}{\begin{equation}}
\newcommand{\ee}{\end{equation}}
\newcommand{\bea}{\begin{eqnarray}}
\newcommand{\eea}{\end{eqnarray}}

\begin{document}
\def\C{{\mathbb{C}}}
\def\R{{\mathbb{R}}}
\def\s{{\mathbb{S}}}
\def\T{{\mathbb{T}}}
\def\Z{{\mathbb{Z}}}
\def\W{{\mathbb{W}}}
\def\Bbb{\mathbb}
\def\BZ{\Bbb Z} \def\BR{\Bbb R}
\def\BW{\Bbb W}
\def\BM{\Bbb M}
\def\e{\mbox{e}}
\def\BC{\Bbb C} \def\BP{\Bbb P}
\def\CP{\BC\BP}
\begin{titlepage}
\title{Generalised Permutation Branes on a product of cosets $G_{k_1}/H\times G_{k_2}/H$}
\author{}
\date{
Gor Sarkissian
\thanks{\noindent E--mail:~ gor.sarkissian@roma2.infn.it}
 \vskip0.4cm
{\sl Dipartimento di Fisica \\
Universita di Roma "Tor Vergata",\\
I.N.F.N sezione di Roma "Tor Vergata"\\
via della Ricerca Scientifica 1, 00133 Roma, Italy}} \maketitle
\abstract{We study the modifications of the generalized
permutation branes defined in hep-th/0509153, which are required
to give rise to the non-factorizable branes on a product of cosets
$G_{k_1}/H\times G_{k_2}/H$. We find that for $k_1\neq k_2$ there
exists big variety of branes, which reduce to the usual
permutation branes, when $k_1=k_2$ and the permutation symmetry is
restored.}
\end{titlepage}

\section{Introduction and summary}\label{intro}
In the recent paper \cite{Fredenhagen:2005an} were suggested the
generalized permutation branes on a product of the WZW models
$G_{k_1}\times G_{k_2}$ with the not necessarily equal levels
$k_1$ and $k_2$. Geometrically the branes wrap the following
submanifolds:
\begin{equation}
\label{geombr} (g_1, g_2)=\{(h_1fh_2^{-1})^{k'_2},\quad
(h_2fh_1^{-1})^{k'_1}|h_1, h_2\in G\}
\end{equation}
where $k'_i=k_i/k$ and $k={\rm gcd}(k_1,k_2)$. Obviously for
$k_1=k_2$ (\ref{geombr}) reduces to the usual permutation branes
\cite{Figueroa-O'Farrill:2000ei,Recknagel:2002qq,Gaberdiel:2002jr,Fuchs:2003yk,Fuchs:1999xn,Fuchs:1999zi,Quella:2002ns}
\begin{equation}
\label{usper} (g_1, g_2)=\{ h_1fh_2^{-1},\quad h_2fh_1^{-1}\}
\end{equation}
 It is well known that for a
submanifold to serve as D-brane in the WZW model the boundary
two-form $\omega_C$ should exist trivializing the  Wess-Zumino
three-form on the brane
\cite{Klimcik:1996hp,Alekseev:1998mc,Gawedzki:1999bq}:
\begin{equation}
\label{cond} \omega^{WZW}|_{brane}=d\omega_{C}\ .
\end{equation}
  It was found in \cite{Fredenhagen:2005an}
that the restriction of the WZW form
$k_1\omega^{WZW}(g_1)+k_2\omega^{WZW}(g_2)$ to the submanifold
(\ref{geombr}) indeed satisfies to this condition with
$\omega_C^{(f)}$ given by the equation:
\begin{eqnarray}
\label{omtwo} \omega^{(f)}_C(h_1,h_2)&=&\frac{k_1k_2}{k}\{{\rm tr}
(h_1^{-1}dh_1fh_2^{-1}dh_2f^{-1})
+{\rm tr}(h_2^{-1}dh_2fh_1^{-1}dh_1f^{-1})\}\nonumber\\
\nonumber
&+&k_1\sum_{j=1}^{k_2'-1}(k_2'-j){\rm tr}(g^j(g^{-1}dg)g^{-j}g^{-1}dg)_{g=h_1fh_2^{-1}}\nonumber \\
&+& k_2\sum_{j=1}^{k_1'-1}(k_1'-j){\rm
tr}(g^j(g^{-1}dg)g^{-j}g^{-1}dg)_{g=h_2fh_1^{-1}}
\end{eqnarray}
From the consideration of the global issues
\cite{Klimcik:1996hp,Alekseev:1998mc,Gawedzki:1999bq} it is
deduced in \cite{Fredenhagen:2005an} that $f=\exp\frac{\pi
i\lambda}{\kappa}$, where $\lambda$ is an integral weight of G Lie
algebra, and $\kappa={\rm lcm}(k_1,k_2)$.

 Comparing the formulae (\ref{geombr}) and (\ref{usper}),
describing generalized and usual permutation branes respectively
one can deduce that the generalized branes preserve less symmetry.
The usual permutation branes preserve two different twisted
adjoint actions:
\begin{equation}
(g_1, g_2)\rightarrow ( mg_1, g_2m^{-1}),\quad (g_1,
g_2)\rightarrow ( g_1m^{-1}, mg_2),
\end{equation}
while the generalized branes preserve only the diagonal subgroup:
\begin{equation}
\label{simb} (g_1, g_2)\rightarrow ( mg_1m^{-1}, mg_2m^{-1})
\end{equation}
Using the form (\ref{omtwo})  boundary equations of motion have
been analyzed  in \cite{Fredenhagen:2005an} and indeed found to be
\begin{equation}
\label{cur} J_1+J_2=\bar{J}_1+\bar{J}_2\, .
\end{equation}

Motivated by the recently established connection between the
permutation D-branes in the minimal $N=2$ supersymmetric models
and the Landau-Ginzburg superpotential factorisation
\cite{Brunner:2005fv,Enger:2005jk,Caviezel:2005th} it was
suggested in \cite{Fredenhagen:2005an} that the generalized
permutation branes should exist also for cosets of the form
$G_{k_1}/H\times G_{k_2}/H$ for the different levels $k_1$ and
$k_2$. To produce such D-brane from the generalized D-branes
(\ref{geombr}) one should modify them in a way to preserve adjoint
action of product $H\times H$:
\begin{equation}
\label{trad} (g_1, g_2)\rightarrow ( m_1g_1m_1^{-1},
m_2g_2m_2^{-1})
\end{equation}
where $m_i\in H$.
  This problem is remained unsolved in
\cite{Fredenhagen:2005an}.

Aim of this paper is to study the modifications of  the ansatz
(\ref{geombr}) giving rise to the required symmetry. In the next
section we show that one can propose two kinds of boundary
conditions possessing with the necessary symmetries.

I.
\begin{equation}
\label{newge} (g_1,
g_2)=\{(h_1fh_2^{-1})^{k_2'}ptp^{-1},\quad\quad
L^{-1}(h_2fh_1^{-1})^{k_1'}nrn^{-1}L\}
\end{equation}
where $L, p, n\in H$ and run all the $H$ subgroup, $r$ and $t$ are
fixed quantized elements of $H$. In other words we multiply both
elements of the ansatz (\ref{geombr}) by the (quantized) conjugacy
classes of the subgroup $H$, and then smear the derived object
along the adjoint action of $H$. The chain of the transformations:
\begin{equation}
\label{grtr} h_1\rightarrow m_1h_1,\quad h_2\rightarrow m_1h_2,
\quad L\rightarrow m_1Lm_2^{-1}, \quad p\rightarrow m_1p, \quad
n\rightarrow m_1n
\end{equation}
reproduces (\ref{trad}).

II.
\begin{equation}
\label{newgee} (g_1,
g_2)=\{(h_1fh_2^{-1})^{k_2'}(s_1ls_2^{-1})^{k_2'},\quad\quad
L^{-1}(h_2fh_1^{-1})^{k_1'}(s_2ls_1^{-1})^{k_1'}L\}
\end{equation}
where $L,s_1, s_2\in H$ run all the subgroup $H$ and $l$ is fixed
quantized element. By words, we multiply the generalized
permutation brane of $G_{k_1}\times G_{k_2}$ by the generalized
permutation brane of $H_{k_3}\times H_{k_4}$ ( $k_3=x_ek_1$,
$k_4=x_ek_2$, $x_e$ is the embedding index of $H$ in $G$
\cite{DiFrancesco:1997nk}) and then again smear derived object
along the adjont action of $H$. We show in the next section that
for  $k_1=k_2$  (\ref{newge}) and (\ref{newgee}) are equivalent.

 Using the Polyakov-Wiegmann identity
\begin{equation}
\omega^{WZW}(gh)=\omega^{WZW}(g)+\omega^{WZW}(h)-d({\rm
tr}(g^{-1}dgdhh^{-1}))
\end{equation}
one obtains that the branes (\ref{newge}) satisfy to the condition
(\ref{cond}) with the following boundary two-form:
\begin{eqnarray}
\label{twnfr}
\Omega(h_1,h_2,L,p,n)&=&\omega_C^{(f)}(h_1,h_2)+k_2\omega^{(2)}(C_2C_4,L)+k_1\omega^{t}(p)
-k_1({\rm tr}(C_1^{-1}dC_1dC_3C_3^{-1}))\nonumber\\ &-&k_2({\rm
tr}(C_2^{-1}dC_2dC_4C_4^{-1}))+k_2\omega^{r}(n)
\end{eqnarray}
where we denoted
\begin{eqnarray}
\label{denot}
C_1&=&(h_1fh_2^{-1})^{k_2'}\, ,\nonumber\\
 C_2&=&(h_2fh_1^{-1})^{k_1'}\, ,\nonumber\\
 C_3&=&ptp^{-1}\, ,\nonumber\\
 C_4&=&nrn^{-1}\, ,
 \end{eqnarray}
$\omega_C^{(f)}$ is the two-form given by the formula
(\ref{omtwo}), $\omega^{t}(p)$ is the two-form found in
\cite{Alekseev:1998mc,Gawedzki:1999bq}
\begin{equation}
\label{omc} \omega^{t}(p)={\rm tr}(p^{-1}dptp^{-1}dpt^{-1})
\end{equation}
and $\omega^{(2)}$ is the following useful two-argument two-form
\be \label{omim} \omega^{(2)}(g,U)={\rm
tr}(dUU^{-1}(gdUU^{-1}g^{-1}+g^{-1}dg+dgg^{-1}))
\end{equation}
which we frequently encounter in this paper.

 For an abelian subgroup $H=U(1)$, and equal levels
$k_1=k_2$ the brane (\ref{newge}) was studied in
\cite{Sarkissian:2003yw}. (In that paper this brane was written in
the form $(h_1fh_2^{-1}L^{-1}, h_2fh_1^{-1}L)$, but after the
redefinition $L^{-1}h_2'=h_2$ we derive (\ref{newge}).) It was
checked for this case in \cite{Sarkissian:2003yw}, that the full
Lagrangian with boundary term (\ref{twnfr}) enjoys with the
symmetry (\ref{trad}).
 It was also shown in \cite{Sarkissian:2003yw} that the geometry of this brane coincides with
the shape corresponding to the permutation boundary state of the
parafermions product. This serves for us as the hint that we found
the correct solution.

One can also check that the branes (\ref{newgee}) as well satisfy
(\ref{cond}) with the two-form
\begin{eqnarray}
\label{twnffr}
\Omega(h_1,h_2,s_1,s_2,L)&=&\omega^{(f)}_C(h_1,h_2)+\omega^{(l)}_C(s_1,s_2)+k_2\omega^{(2)}(C_2C_6,L)
-k_1({\rm tr}(C_1^{-1}dC_1dC_5C_5^{-1}))\nonumber\\ &-&k_2({\rm
tr}(C_2^{-1}dC_2dC_6C_6^{-1}))
\end{eqnarray}
where
\begin{eqnarray}
\label{denot2}
C_5&=&(s_1ls_2^{-1})^{k_2'}\, ,\nonumber\\
 C_6&=&(s_2ls_1^{-1})^{k_1'}\, .
 \end{eqnarray}
 The rest of the paper is organized as
follows.

In the section 2 we obtain boundary conditions (\ref{newge}) and
(\ref{newgee}) by analyzing the action of the gauged WZW model on
a world-sheet with boundary.

In the section 3 we consider these branes  for product
$SU(2)_{k_1}/U(1)\times SU(2)_{k_2}/U(1)$.

In the appendix A we deliver some algebraical calculations proving
gauge invariance of action with boundary term given by
(\ref{twnfr}).

In the appendix B we review different coordinates systems on $S^3$
sphere used in the section 3.

\section{Lagrangian and symmetries}
In this section we obtain the boundary conditions (\ref{newge})
and (\ref{newgee}) by considering gauged WZW model action on a
world-sheet with boundary along the way worked out in
\cite{Elitzur:2001qd} and \cite{Kubota:2001ai}. First of all we
remind the bulk action of the gauged WZW model
\cite{Bardakci:1987ee,Gawedzki:1988hq,Gawedzki:1988nj,Karabali:1988au,Karabali:1989dk}:
\begin{equation}
\label{gact}
S^{G/H}(g,A)=S^{G}+\frac{k_G}{2\pi}\int_{\Sigma}d^2z{\rm
tr}\{A_{\bar{z}}\partial_zgg^{-1}-A_zg^{-1}\partial_{\bar{z}}g+A_{\bar{z}}gA_zg^{-1}-A_zA_{\bar{z}}\}
\end{equation}
where
\begin{equation}
S^G=\frac{k_G}{4\pi}\left(\int_{\Sigma}d^2zL^{{\rm
kin}}+\int_B\omega^{WZW}(g)\right)
\end{equation}
is bulk WZW action, $L^{{\rm kin}}={\rm
tr}(\partial_zg\partial_{\bar{z}}g^{-1})$,
$\omega^{WZW}=\frac{1}{3}{\rm tr}(dgg^{-1})^3$, $B$ is a
three-dimensional manifold bounded by $\Sigma$, and $A$ is a gauge
field taking values in the $H$ Lie algebra.

Defining
\begin{equation}
A_z=\partial_z\tilde{U}\tilde{U}^{-1},\quad
A_{\bar{z}}=\partial_zUU^{-1}
\end{equation}
we can write (\ref{gact}) in the form:
\begin{equation}
\label{ginvac}
S^{G/H}(\tilde{g},\tilde{h})=S^{G}(\tilde{g})-S^H(\tilde{h})
\end{equation}
where $\tilde{g}=U^{-1}g\tilde{U}$ and
$\tilde{h}=U^{-1}\tilde{U}$. The level $k_H$ of the $S^{H}$ is
related to $k_G$ through the embedding index $x_e$ of $H$ in $G$:
 $k_H=x_ek_G$ \cite{DiFrancesco:1997nk}. The expression
(\ref{ginvac}) is manifestly gauge invariant under the
transformation: \be g\rightarrow mgm^{-1}, \quad U\rightarrow
mU,\quad \tilde{U}\rightarrow m\tilde{U}\ee
 For the case under consideration the bulk action is:
\begin{equation}
\label{dueact} S_{\rm
bulk}^{G/H}(\tilde{g}_1,\tilde{g}_2,\tilde{h}_1,\tilde{h}_2)=S^{G}(\tilde{g}_1)+S^{G}(\tilde{g}_2)-S^H(\tilde{h}_1)-S^{H}(\tilde{h}_2)
\end{equation}
where
\begin{eqnarray}
\tilde{g}_1&=&U^{-1}_1g_1\tilde{U}_1\nonumber\\
 \tilde{g}_2&=&U^{-1}_2g_2\tilde{U}_2\nonumber\\
\tilde{h}_1&=&U^{-1}_1\tilde{U}_1\nonumber\\
 \tilde{h}_2&=&U^{-1}_2\tilde{U}_2
\end{eqnarray}
The levels $k_3$ and $k_4$ of the third and forth terms in
(\ref{dueact}), as explained above, are related to the levels
$k_1$ and $k_2$ of the first and second terms respectively through
the embedding index $x_e$ of $H$ in $G$  \bea k_3&=&x_ek_1\nonumber\\
k_4&=&x_ek_2\, .\eea
 Consider the action (\ref{dueact}) in the presence of
boundary. We should specify boundary conditions. Let us choose for
the sum of the first two actions the generalized permutation
boundary conditions (\ref{geombr}) and impose the arguments of the
third and forth terms to take their values in the quantized
conjugacy classes:
\begin{eqnarray}
\label{bndcn}
\tilde{g}_1&=&((U_1^{-1}h_1)f(U_1^{-1}h_2)^{-1})^{k_2'}\nonumber\\
\tilde{h}_1&=&(U_1^{-1}p)t^{-1}(U_1^{-1}p)^{-1}\nonumber\\
\tilde{g}_2&=&((U_1^{-1}h_2)f(U_1^{-1}h_1)^{-1})^{k_1'}\nonumber\\
\tilde{h}_2&=&(U_1^{-1}n)r^{-1}(U_1^{-1}n)^{-1}
\end{eqnarray}
 It is
easy to see that (\ref{bndcn}) brings to us conditions
(\ref{newge}) for $g_1$ and $g_2$ with
\begin{equation}
\label{ldef}
 L=U_1U_2^{-1}
\end{equation}
  Using the
forms (\ref{omtwo}) and (\ref{omc}) one can write the full action
with the boundary term:
\begin{equation}
\label{acc} S=S_{\rm
bulk}^{G/H}(\tilde{g}_1,\tilde{g}_2,\tilde{h}_1,\tilde{h}_2)-\frac{1}{4\pi}\int_D(\omega^{(f)}_C(U_1^{-1}h_1,U_1^{-1}h_2)-
k_1\omega^{(t^{-1})}(U_1^{-1}p)-k_2\omega^{(r^{-1})}(U_1^{-1}n)
\end{equation}
where $\partial B=\Sigma +D$.

 The action (\ref{acc}) is manifestly
gauge invariant under the gauge transformation:
\begin{equation}
h_1\rightarrow m_1h_1,\quad h_2\rightarrow m_1h_2, \quad
U_1\rightarrow m_1U_1,\quad U_2\rightarrow m_2U_2, \quad
p\rightarrow m_1p, \quad n\rightarrow m_1n
\end{equation}
implying, recalling the definition (\ref{ldef}), \be L\rightarrow
m_1Lm_2^{-1}\, .\ee We derived the transformation rules
(\ref{grtr}).
  Using
the Polyakov-Wiegmann identities the action (\ref{acc}) can be
written as :
\begin{equation}
S=S^{G/H}(g_1,A_1)+S^{G/H}(g_2,A_2)-\frac{1}{4\pi}\int_D \Omega
\end{equation}
where
\begin{equation}
\label{monfr}
\Omega=\omega^{(f)}_C(U_1^{-1}h_1,U_1^{-1}h_2)+k_1\omega^{(t)}(U_1^{-1}p)+k_2\omega^{(r)}(U_1^{-1}n)+k_1\omega(g_1,U_1,\tilde{U}_1)+k_2\omega(g_2,U_2,\tilde{U}_2)
\end{equation}

where \be\label{ompw} \omega(g_i,U_i,\tilde{U}_i)={\rm
tr}(g^{-1}_idg_id\tilde{U}_i\tilde{U}_i^{-1}-dU_iU_i^{-1}dg_ig_i^{-1}-dU_iU_i^{-1}g_id\tilde{U}_i\tilde{U}_i^{-1}g_i^{-1}+
dU_iU_i^{-1}d\tilde{U}_i\tilde{U}_i^{-1})\ee
 It is cumbersome but straightforward to check that the form
(\ref{monfr}) coincides with (\ref{twnfr}) with $L$ given by
(\ref{ldef}). The details of the calculations  are delivered in
the appendix A.

Boundary conditions (\ref{newgee}) can be received in the same
way, but now one should take the generalized boundary conditions
as for the sum of the first two actions, as well for the sum of
actions for the gauge groups:
\begin{eqnarray}
\label{bndcnn}
\tilde{g}_1&=&((U_1^{-1}h_1)f(U_1^{-1}h_2)^{-1})^{k_2'}\nonumber\\
\tilde{h}_1&=&((U_1^{-1}s_2)l^{-1}(U_1^{-1}s_1)^{-1})^{k_2'}\nonumber\\
\tilde{g}_2&=&((U_1^{-1}h_2)f(U_1^{-1}h_1)^{-1})^{k_1'}\nonumber\\
\tilde{h}_2&=&((U_1^{-1}s_1)l^{-1}(U_1^{-1}s_2)^{-1})^{k_1'}
\end{eqnarray}
where we have taken into account that $\tilde{k}={\rm
gcd}(k_3,k_4)=x_e{\rm gcd}(k_1,k_2)$, and
$k_3'=k_3/\tilde{k}=k_1'$, $k_4'=k_4/\tilde{k}=k_2'$.

 One can easily check
that these conditions are equivalent to (\ref{newgee}) with again
$L=U_1U_2^{-1}$.

The full action with boundary term is:
\begin{equation}
\label{acc2} S=S_{\rm
bulk}^{G/H}(\tilde{g}_1,\tilde{g}_2,\tilde{h}_1,\tilde{h}_2)-\frac{1}{4\pi}\int_D(\omega^{(f)}_C(U_1^{-1}h_1,U_1^{-1}h_2)-
\omega^{(l^{-1})}_C(U_1^{-1}s_2,U_1^{-1}s_1)
\end{equation}
where $\partial B=\Sigma +D$. The action (\ref{acc2}) is
manifestly gauge invariant under the gauge transformation:
\begin{equation}
h_1\rightarrow m_1h_1,\quad h_2\rightarrow m_1h_2, \quad
U_1\rightarrow m_1U_1,\quad U_2\rightarrow m_2U_2, \quad
s_1\rightarrow m_1s_1, \quad s_2\rightarrow m_1s_2
\end{equation}
implying, again, \be L\rightarrow m_1Lm_2^{-1}\, .\ee
 Using
Polyakov-Wiegmann identities the action (\ref{acc2}) can be
written as :
\begin{equation}
S=S^{G/H}(g_1,A_1)+S^{G/H}(g_2,A_2)-\frac{1}{4\pi}\int_D \Omega
\end{equation}
where
\begin{equation}
\label{monfr2}
\Omega=\omega^{(f)}_C(U_1^{-1}h_1,U_1^{-1}h_2)+\omega^{(l)}_C(U_1^{-1}s_1,U_1^{-1}s_2)+k_1\omega(g_1,U_1,\tilde{U}_1)+k_2\omega(g_2,U_2,\tilde{U}_2)
\end{equation}
Repeating the same steps as outlined in the appendix A one can
show that (\ref{monfr2}) coincides with (\ref{twnffr}).

Some comments:

Let us consider the branes (\ref{newge}) and (\ref{newgee}) for
equal levels $k_1=k_2$, when permutation symmetry is restored: \be
\label{prbr1}
(g_1,g_2)=(C_1C_3,L^{-1}C_2C_4L)=(h_1fh_2^{-1}ptp^{-1},L^{-1}h_2fh_1^{-1}nrn^{-1}L)\ee
\be \label{prbr2}
(g_1,g_2)=(C_1C_5,L^{-1}C_2C_6L)=(h_1fh_2^{-1}s_1ls_2^{-1},L^{-1}h_2fh_1^{-1}s_2ls_1^{-1}L)\ee

  By the redefinition \be \label{redef} h_2^{-1}C_5=h'_2,\quad
L'=C_5^{-1}L\ee one can write the brane (\ref{prbr2}) in the form
$(C_1, L^{-1}C_2C_6C_5L)$. Taking into account that $C_6C_5$  in
this case is the usual conjugacy class, we see that at the point
of the restored symmetry the family of the branes (\ref{prbr2})
coincides with the family (\ref{prbr1}).

Performing the same kind of redefinition in (\ref{prbr1}) \be
\label{redef2} h_2^{-1}C_3=h'_2,\quad L'=C_3^{-1}L\ee one can
write all the branes (\ref{prbr1}) in the form \be \label{prbr3}
(g_1,g_2)=(C_1,L^{-1}C_2C_4C_3L)\ee

Presumably we can multiply both elements in (\ref{newge}) by the
chain of conjugacy classes, as in \cite{Quella:2002ns}, but for
the case of $k_1=k_2$, when permutation symmetry is restored, we
see, that one can cover all the family already multiplying just
one of them. The conclusion is, that generically when $k_1\neq
k_2$ we have two families of branes (\ref{newge}) and
(\ref{newgee}) , which reduce to the branes of the form
(\ref{prbr3}), when the permutation symmetry is restored.

\section{Generalized permutation branes on $SU(2)_{k_1}/U(1)\times SU(2)_{k_2}/U(1)$}

In this section we consider permutation branes on product of
$SU(2)_{k_1}/U(1)\times SU(2)_{k_2}/U(1)$ cosets. At the beginning
we consider usual permutation brane for $k_1=k_2$ and show that
the geometrical description given above coincide with the
permutation boundary state \cite{Recknagel:2002qq} overlap with
the graviton wave packet. Actually this calculation was performed
in \cite{Sarkissian:2003yw}, but for the completeness and the
reader's convenience we repeat it (slightly generalized and with
corrected typos) here. Then we elaborate the geometry of the
simplest generalized permutation brane.

With the $U(1)$ subgroup generated by $\sigma_3$ the brane
(\ref{newge}) for $k_1=k_2=k$ takes the form:
\begin{equation}
\label{nnbrd} (g_1,g_2)\Big|_{{\rm brane}}=\big(h_1fh_2^{-1},\,\,
e^{i\alpha{\sigma_3\over 2}}h_2h_1^{-1}e^{-i\alpha{\sigma_3\over
2}}e^{i\frac{\pi M}{k}\frac{\sigma_3}{2}}\big) \, .
\end{equation}
where $f=e^{i\hat{\psi}\frac{\sigma_3}{2}}$,
  $\hat{\psi}=\frac{2j\pi}{k}$, $j=0,\ldots,\frac{k}{2}$,  and $M$  is an integer. The factor
$e^{i\frac{\pi M}{k}\frac{\sigma_3}{2}}$ reflects ${\cal Z}_k$
symmetry of an abelian coset \cite{Maldacena:2001ky}. One can
multiply with this factor also the first element in (\ref{nnbrd})
, but as explained above, performing the redefinition
(\ref{redef2}), one gets again (\ref{nnbrd}). We see that all the
branes are labelled by two indices $\hat{\psi}$ and $M$, exactly
as the permutation states of the parafermions product.
 The elements $g_1$ and $g_2$ belong to the brane surface if
the following equation admits a solution for the parameter
$\alpha$,
\begin{equation}
\label{nnbrddd} {\rm tr}\left(g_1e^{-i\alpha{\sigma_3\over
2}}g_2e^{i\alpha{\sigma_3\over 2}}e^{-i\frac{\pi
M}{k}\frac{\sigma_3}{2}}\right)=2\cos\hat{\psi} \, .
\end{equation}
This equation can be further elaborated in the Euler coordinates,
reviewed in the appendix B.
 The formulae for the Euler angles of a product of two
elements $\hat{g}=g_1g_2$ are given in~\cite{Vilenkin:1968nk}

\begin{equation}
\label{thetpr} \cos \hat{\tilde{\theta}} =
\cos\tilde{\theta}_1\cos\tilde{\theta}_2-\sin\tilde{\theta}_1\sin\tilde{\theta}_2
\cos(\chi_2+\varphi_1) \, ,
\end{equation}
\begin{equation}
 \label{phipr} e^{i\hat{\tilde{\phi}}}
= {e^{i{\chi_1+\varphi_2\over 2}} \over \cos
{\hat{\tilde{\theta}}\over 2} } \left( \cos{\tilde{\theta}_1\over
2} \cos{\tilde{\theta}_2\over 2}e^{i {\chi_2+\varphi_1\over 2}}-
\sin{\tilde{\theta}_1\over 2}\sin{\tilde{\theta}_2\over 2} e^{-i
{\chi_2+\varphi_1\over 2}} \right) \, .
\end{equation}

where the hatted variables refer to the product $\hat{g}$.

Denoting by $\tilde{\Theta}$, $\tilde{\Phi}$ Euler angles
$\tilde{\theta}$ and $\tilde{\phi}$ of the product
$g_1e^{-i\alpha\frac{\sigma_3}{2}}g_2$ and  using (\ref{thetpr})
and (\ref{phipr}) we can rewrite (\ref{nnbrddd}) as
\begin{equation}
\label{baseq1} \cos{\tilde{\Theta}\over 2}
\cos(\gamma/2-\xi/2-\tilde{\phi}_1-\tilde{\phi}_2+\frac{\pi
M}{2k})=\cos\hat{\psi}\,,
\end{equation}
where
\begin{equation}
\label{thhh}
\cos\tilde{\Theta}=\cos\tilde{\theta}_1\cos\tilde{\theta}_2-\sin\tilde{\theta}_1\sin\tilde{\theta}_2\cos\gamma\,
,
\end{equation}
and we have introduced new labels $\gamma=\chi_2+\varphi_1-\alpha$
and $\xi/2=\tilde{\Phi}-{\chi_1+\varphi_2\over 2}$. The variables
$\xi$ and $\gamma$ are related to each other by the equation
\begin{equation}
\label{phiprrr} e^{i{\xi\over 2}}={ 1 \over
\cos{\tilde{\Theta}\over 2}} \left( \cos{\tilde{\theta}_1\over
2}\cos{\tilde{\theta}_2\over 2}e^{i{\gamma \over 2}}-
\sin{\tilde{\theta}_1\over 2}\sin{\tilde{\theta}_2\over
2}e^{-i{\gamma\over 2}} \right) \, .
\end{equation}

Let us recall that the vectorial gauging of $U(1)$ symmetry is
corresponding to the translation of  $\phi$ and the resulting
target space of the $SU(2)_k/U(1)$ model, derived after the gauge
fixing $\phi=0$ and integrating out of the gauge field, is the
two-dimensional disc,
 parameterized by $\theta$ and $\tilde{\phi}$.
In the case of product the target space is parameterized by
$\theta_1, \theta_2,\tilde{\phi}_1,\tilde{\phi}_2$. Hence the
brane consists of those points for which equation~(\ref{baseq1})
admits a solution for~$\gamma$. $\Theta$ and $\xi$ are considered
here as the complicated functions of
$\tilde{\theta_1},\tilde{\theta_2}$ and $\gamma$ given by
(\ref{thhh}) and (\ref{phiprrr}) respectively.
 For~$\hat{\psi}=0$
there are additional constraints, which imply that in this case
the brane is two dimensional and given by the equations
\begin{equation}
\label{nnbrdd} \tilde{\theta}_1=-\tilde{\theta}_2, \quad
\tilde{\phi}_1=-\tilde{\phi}_2 +\frac{\pi M}{2k}\, .
\end{equation}
Now we calculate the effective geometry corresponding to the
permutation boundary state \cite{Recknagel:2002qq}:
\begin{equation}
|L,M\rangle=\sum_{j,m}\frac{S_{Lj}}{S_{0j}}e^{i\pi
Mm/k}\sum_{N_1,N_2}|j,m,N_1\rangle_1\otimes\overline{|j,m,N_1\rangle_2}\otimes|j,m,N_2\rangle_2\otimes\overline{|j,m,N_2\rangle_1}
\end{equation}
where $S_{Lj}$ is matrix of the modular transformation of
$SU(2)_k$
\begin{equation}
\label{smatr} S_{Lj}=\sqrt{{2\over
k+2}}\sin\left({(2L+1)(2j+1)\pi\over k+2}\right) \, .
\end{equation}

 To obtain the effective
geometry, one should compute the overlap
$\langle\theta_1,\tilde{\phi}_1,\theta_2,\tilde{\phi}_2|L,M\rangle$.
At the beginning we should find the wave-functions of the
parafermion disc theory \cite{Maldacena:2001ky}:
\begin{equation}
\Psi_{j,m}(\theta,\tilde{\phi})=\langle\theta,\tilde{\phi}|j,m\rangle\rangle
\end{equation}
The wave-functions of the disc are the $SU(2)$ wave-functions that
are invariant under translation of $\phi$. (Note that in
 \cite{Maldacena:2001ky} axial gauging is considered, and as a
 consequence
 the roles of $\phi$ and $\tilde{\phi}$ are interchanged).
Recalling that the $SU(2)$ wave-functions are the normalized
Wigner functions
\begin{equation}
\sqrt{2j+1}{\cal
D}^j_{nm}(g(\vec{\theta}))=\sqrt{2j+1}e^{-i(n\chi+m\varphi)}d^j_{nm}(\cos\tilde{\theta})
\, ,
\end{equation}
we see that the function on disc are those of them with $m=n$.
Using that for the large $k$
\begin{equation}
\label{lark} {S_{Lj}\over S_{0j}}\sim {(k+2)\over \pi
(2j+1)}\sin[(2j+1)\hat{\psi}] \, ,
\end{equation}
where $\hat{\psi}=\frac{(2L+1)\pi}{k+2}$, one obtains that
 in the large-$k$ limit the overlap
reduces to
\begin{equation}
\langle\vec{\theta}_1,\vec{\theta}_2|L,M\rangle\sim
\sum_j\sum_m\sin[(2j+1)\hat{\psi}]e^{i\pi Mm/k}{\cal
D}^j_{mm}(g_1(\vec{\theta}_1)){\cal D}^j_{mm}(g_2(\vec{\theta}_2))
\, .
\end{equation}

 It is known \cite{Vilenkin:1968nk} that $d_{nm}^j$ are satisfying the relation
 (note that there is no summation assumed for the repeated
indices)
\begin{equation}
\label{dprod}
d^j_{mm}(\cos\tilde{\theta}_1)d^j_{mm}(\cos\tilde{\theta}_2)={1\over
2\pi}\int_{-\pi}^{\pi}
e^{im(\gamma-\xi)}d^j_{mm}(\cos\tilde{\Theta})d\gamma \, ,
\end{equation}
The functions $\tilde{\Theta}$ and $\xi$ are functions of
$\tilde{\theta}_1,\tilde{\theta}_2$ and $\gamma$ defined in
equations~(\ref{thhh}) and (\ref{phiprrr}). Using (\ref{dprod})
the overlap of the boundary state with the bulk probe can be
written as
\begin{equation}
\langle\vec{\theta}_1,\vec{\theta}_2|L,M\rangle \sim
\sum_j\sum_m\int_{-\pi}^{\pi}\sin[(2j+1)\hat{\psi}]
e^{im(\gamma-\xi-2\tilde{\phi}_1-2\tilde{\phi}_2+\frac{\pi
M}{k})}d^j_{mm}(\cos\tilde{\Theta})d\gamma
\end{equation}
Now using that $\sum_m {\cal
D}^j_{mm}(g)=\frac{\sin(2j+1)\psi}{\sin\psi}$, where $\psi$ is the
angle of the standard metric (\ref{standard-s3}) and defined by
the relation ${\rm Tr}g = 2 \cos \psi$, and the completeness of
$\sin[(2j+1)\psi]$ on the interval $[0,\pi]$ we get
\begin{equation}
\langle\vec{\theta}_1,\vec{\theta}_2|L,M\rangle\sim
\int_{-\pi}^{\pi}{\delta(\psi-\hat{\psi})\over
\sin\hat{\psi}}d\gamma \, ,
\end{equation}
where
\begin{equation}
\cos\psi=\cos{\tilde{\Theta}\over 2}
\cos(\gamma/2-\xi/2-\tilde{\phi}_1-\tilde{\phi}_2+\frac{\pi
M}{2k})
\end{equation}
From this equation it follows that the brane consist of all those
points for which the expression in the argument of the~$\delta$
function has a root for~$\gamma$. This is the same condition as
the one coming from equation~(\ref{baseq1}), obtained in the
Langrangian approach.

Let us turn now to the generalized permutation brane on a product
$SU(2)_{k_1}/U(1)\times SU(2)_{k_2}/U(1)$. The branes
(\ref{newge}) and (\ref{newgee}) for abelian case take forms
\begin{equation}
\label{newgab1} {\cal H}=\{(h_1fh_2^{-1})^{k_2'}e^{i\frac{\pi
M_1}{k_1}\frac{\sigma_3}{2}},\quad e^{i\alpha{\sigma_3\over
2}}(h_2fh_1^{-1})^{k_1'}e^{-i\alpha{\sigma_3\over 2}}e^{i\frac{\pi
M_2}{k_2}\frac{\sigma_3}{2}}\}
\end{equation}
\begin{equation}
\label{newgab2} {\cal H}=\{(h_1fh_2^{-1})^{k_2'}e^{i\beta
k_2'{\sigma_3\over 2}}e^{i\frac{\pi
M_1}{k_1}\frac{\sigma_3}{2}},\quad e^{i\alpha{\sigma_3\over
2}}(h_2fh_1^{-1})^{k_1'}e^{-i\beta k_1'{\sigma_3\over
2}}e^{-i\alpha{\sigma_3\over 2}}e^{i\frac{\pi
M_2}{k_2}\frac{\sigma_3}{2}}\}
\end{equation}
We see that for $k_1\neq k_2$ we have much bigger variety of
branes, which all degenerate to (\ref{nnbrd}) when $k_1=k_2$.

  Now we describe geometry of the generalized permutation
branes (\ref{newgab1}) and (\ref{newgab2}) for the simplest case
$f=e$. For this case the branes are:
\begin{equation}
\label{simbr} (g_1,g_2)=(g^{k_2'}e^{i\frac{\pi
M_1}{k_1}\frac{\sigma_3}{2}},\quad e^{i\alpha{\sigma_3\over
2}}g^{-k_1'}e^{-i\alpha{\sigma_3\over 2}}e^{i\frac{\pi
M_2}{k_2}\frac{\sigma_3}{2}})
\end{equation}
\begin{equation}
\label{simbr2} (g_1,g_2)=(g^{k_2'}e^{i\beta k_2'{\sigma_3\over
2}}e^{i\frac{\pi M_1}{k_1}\frac{\sigma_3}{2}},\quad
e^{i\alpha{\sigma_3\over 2}}g^{-k_1'}e^{-i\alpha{\sigma_3\over
2}}e^{-i\beta k_1'{\sigma_3\over 2}}e^{i\frac{\pi
M_2}{k_2}\frac{\sigma_3}{2}})
\end{equation}
To elaborate the geometry of (\ref{simbr}) and (\ref{simbr2}) we
first recall the useful fact mentioned in
\cite{Fredenhagen:2005an} that the element $g^n$ in the
coordinates (\ref{standard-s3}) has the same angles $\xi$ and
$\eta$ as $g$ but $\psi$ has to be replaced by $n\psi$. Then we
need the formulae of transformation from the coordinates
(\ref{standard-s3}) to the Euler coordinates:
\begin{eqnarray}
\tan\tilde{\phi}&=&\tan\psi\cos\xi\nonumber\\
\sin\theta&=& \sin\psi\sin\xi
\end{eqnarray}
Taking finally into account that the adjoint action of $U(1)$ does
not change $\theta$ and $\tilde{\phi}$ angles of an element one
can describe the geometry of (\ref{simbr}) and (\ref{simbr2}) by
the following equations of embedding respectively:
\begin{eqnarray}
\label{simp}
 \tan\left(\tilde{\phi}_1-\frac{\pi
M_1}{2k_1}\right)&=&\tan k_2'\psi\cos\xi\nonumber\\
\sin\theta_1&=& \sin k_2'\psi\sin\xi\nonumber\\
\tan\left(\tilde{\phi}_2-\frac{\pi
M_2}{2k_2}\right)&=&-\tan k_1'\psi\cos\xi\nonumber\\
\sin\theta_2&=& -\sin k_1'\psi\sin\xi
\end{eqnarray}
\begin{eqnarray}
\label{simp2}
 \tan\left(\tilde{\phi}_1-k_2'\beta-\frac{\pi
M_1}{2k_1}\right)&=&\tan k_2'\psi\cos\xi\nonumber\\
\sin\theta_1&=& \sin k_2'\psi\sin\xi\nonumber\\
\tan\left(\tilde{\phi}_2+k_1'\beta-\frac{\pi
M_2}{2k_2}\right)&=&-\tan k_1'\psi\cos\xi\nonumber\\
\sin\theta_2&=& -\sin k_1'\psi\sin\xi
\end{eqnarray}
The brane (\ref{simbr}) is two-dimensional with the world-volume
coordinates $(\psi,\xi)$, whereas the brane (\ref{simbr2}) is
three-dimensional with the world-volume coordinates
$(\psi,\xi,\beta)$. It is easy to check that when $k_1=k_2$, and
$k_1'=k_2'=1$, (\ref{simp}) and (\ref{simp2}) reduce to
(\ref{nnbrdd}).
 To find various other properties of the branes like mass,
spectrum {\it et.c.}, is left for the future work.
\newpage

\appendix

\section{Details of calculations}

 To show that (\ref{monfr}) coincides with (\ref{twnfr}) at the beginning  we solve (\ref{bndcn}) :
\begin{equation}
\label{c1} \tilde{U}_1=C_3^{-1}U_1,
\end{equation}
\begin{equation}
\label{cc1} \tilde{U}_2=U_2U_1^{-1}C_4^{-1}U_1,
\end{equation}
\begin{equation}
\label{c2}
 g_1=C_1C_3
 \end{equation}
 \begin{equation}
 \label{c3}
g_2=L^{-1}C_2C_4L=(U_1U_2^{-1})^{-1}C_2C_4(U_1U_2^{-1})
\end{equation}
where $C_1$, $C_2$, $C_3$, $C_4$ are defined in (\ref{denot}).

Inserting (\ref{c1}) and (\ref{c2}) in (\ref{ompw}) for $i=1$ one
can show that
 \begin{equation}
 \label{secli}
 \omega(g_1,U_1,\tilde{U}_1)=-{\rm
 tr}(C_1^{-1}dC_1dC_3C_3^{-1})-\omega^{(2)}(C_1,U_1)-\omega^{(2)}(C_3,U_1)
\end{equation}
 Inserting  (\ref{cc1}) and (\ref{c3}) in (\ref{ompw}) for $i=2$ one obtains
\begin{equation}
\label{o2}
\omega^{(2)}(g_2,U_2,\tilde{U}_2)=-\omega^{(2)}(C_2,U_1)-\omega^{(2)}(C_4,U_1)+\omega^{(2)}(C_2C_4,U_1U_2^{-1})-{\rm
 tr}(C_2^{-1}dC_2dC_4C_4^{-1})
\end{equation}

 To deal with the second and third terms in (\ref{monfr}) we recall the following useful identity derived in
\cite{Elitzur:2001qd}:
\begin{equation}
\label{usid}
\omega^{(t)}(U_1^{-1}p)=\omega^{(t)}(p)+\omega^{(2)}(C_3,U_1)
\end{equation}
It was shown in \cite{Elitzur:2001qd} that (\ref{usid}) guarantees
that the full WZW Lagrangian on a world-sheet with boundary, with
boundary conditions specified  by the conjugacy class $C_3$,
enjoys with the full diagonal subalgebra: \be
g(z,\bar{z})\rightarrow k_L(z)g(z,\bar{z})k^{-1}_R(\bar{z}),\quad
k_L|_{\rm boundary}=k_R|_{\rm boundary}\ee

 The last ingredient which we need is the formula giving
transformation properties of $\omega_C$:
\begin{equation}
\label{trpr}
\omega^{(f)}_C(U_1^{-1}h_1,U_1^{-1}h_2)=\omega^{(f)}_C(h_1,h_2)+k_1\omega^{(2)}(C_1,U_1)+k_2\omega^{(2)}(C_2,U_1)
\end{equation}
It is possible to derive this formula using the definition
(\ref{omtwo}) . But this formula is nothing else as the global
form of the equation (\ref{cur}), reflecting symmetry properties
of the generalized permutation brane. It is straightforward to
check that (\ref{trpr}) guarantees that the WZW Lagrangian on a
world-sheet with boundary with the boundary conditions given by
the generalized permutation brane (\ref{geombr}) enjoys with the
symmetry \bea g_1(z,\bar{z})&\rightarrow
&k_L(z)g_1(z,\bar{z})k^{-1}_R(\bar{z}), \quad
g_2(z,\bar{z})\rightarrow
h_L(z)g_2(z,\bar{z})h^{-1}_R(\bar{z}),\nonumber\\ k_L|_{\rm
boundary}&=&k_R|_{\rm boundary}=h_L|_{\rm boundary}=h_R|_{\rm
boundary}\eea
 Inserting (\ref{secli}), (\ref{o2}),
(\ref{usid}) and (\ref{trpr}) and in (\ref{monfr}) one ends up
with (\ref{twnfr}).

\section{Various coordinate systems for the sphere and relations between them}

A three-sphere $S^3$ is a group manifold of the $SU(2)$ group. A
generic element in this group can be written as
\begin{equation}
\label{SU(2)} g =  \, X_0 \sigma_0 + i (X_1 \sigma_1 + X_2
\sigma_2 + X_3 \sigma_3) = \pmatrix{X_0 + i X_3 \, & \, X_2 + iX_1
\cr -(X_2 - i X_1)  \, & \, X_0 - iX_3\cr}
\end{equation}
subject to condition that the determinant  is equal to one
\begin{equation}
\label{hyper} X_{0}^2 + X_1^2 + X_2^2+X_{3}^2 = 1 \, .
\end{equation}
The metric on $S^3$ can be written in the following three ways,
which will be used in the main text. Firstly, using the Euler
parametrization of the group element we have
\begin{eqnarray}
\label{euler-s3}
g &=& e^{i\chi {\sigma_3 \over 2}}e^{i \tilde{\theta} {\sigma_1 \over 2}} e^{i \varphi {\sigma_3 \over 2}} \, \\
{\rm d}s^2 &=& {1\over 4} \Big( ({\rm d} \chi + \cos
\tilde{\theta} {\rm d} \varphi)^2 + {\rm d} \tilde{\theta}^2 +
\sin^2 \tilde{\theta} {\rm d} \varphi^2 \Big) \, .
\end{eqnarray}
The ranges of coordinates are $0 \leq \tilde{\theta}\, \leq \pi$,
$0 \leq \varphi \leq 2 \pi$ and $0 \leq \chi \leq 4 \pi$.

Secondly, we can use coordinates that are analogue to the global
coordinate for~$AdS_3$
\begin{eqnarray}
\label{global-s3}
X_0+iX_3 &=& \cos \theta e^{i\tilde{\phi}} \, , \quad X_2+iX_1 = \sin \theta e^{i\phi}\, \\
{\rm d}s^2 &=& {\rm d} \theta^2 + \cos^2 \theta {\rm d}
\tilde{\phi}^2 + \sin^2 \theta {\rm d} \phi^2 \, .
\end{eqnarray}
The relation between the metrics~(\ref{euler-s3})
and~(\ref{global-s3}) is given by
\begin{equation}
\label{glob-par} \chi = \tilde{\phi} + \phi  \, , \quad \varphi =
\tilde{\phi} - \phi \, ,
 \quad \theta={\tilde{\theta} \over 2} \,.
\end{equation}
The ranges of coordinates are $-\pi\leq\tilde{\phi},\phi\leq\pi$
and $0\leq\theta\leq{\pi\over 2}$.

Thirdly, the standard metric on~$S^3$ is given by ($\vec{n}$ is a
unit vector on~$S^2$)
\begin{eqnarray}
\label{standard-s3} g &=& e^{2i  \psi {\vec{n} \cdot \vec{\sigma}
\over 2}} \, , \quad
{\rm d}s^2 = {\rm d} \psi^2 + \sin^2 \psi ( {\rm d} \xi^2 + \sin^2 \xi {\rm d} \eta^2) \, \\
X_0+iX_3&=& \cos \psi +i\sin \psi\cos \xi \, , \quad X_2+iX_1 =
\sin \psi \sin \xi e^{i\eta} \,.
\end{eqnarray}
The ranges of the coordinates are $0 \leq\psi\, ,\, \xi  \leq \pi$
and $ 0\leq \eta \leq  2 \pi$.


\newpage
\begin{thebibliography}{99}
\bibitem{Fredenhagen:2005an}
  S.~Fredenhagen and T.~Quella,
  ``Generalised permutation branes,''
  JHEP {\bf 0511} (2005) 004
  [arXiv:hep-th/0509153].
\bibitem{Figueroa-O'Farrill:2000ei}
  J.~M.~Figueroa-O'Farrill and S.~Stanciu,
  ``D-branes in AdS(3) x S(3) x S(3) x S(1),''
  JHEP {\bf 0004} (2000) 005
  [arXiv:hep-th/0001199].
\bibitem{Recknagel:2002qq}
  A.~Recknagel,
  ``Permutation branes,''
  JHEP {\bf 0304} (2003) 041
  [arXiv:hep-th/0208119].
\bibitem{Gaberdiel:2002jr}
  M.~R.~Gaberdiel and S.~Schafer-Nameki,
  ``D-branes in an asymmetric orbifold,''
  Nucl.\ Phys.\ B {\bf 654} (2003) 177
  [arXiv:hep-th/0210137].
\bibitem{Fuchs:2003yk}
  J.~Fuchs, I.~Runkel and C.~Schweigert,
  ``Boundaries, defects and Frobenius algebras,''
  Fortsch.\ Phys.\  {\bf 51} (2003) 850
  [Annales Henri Poincare {\bf 4} (2003) S175]
  [arXiv:hep-th/0302200].
\bibitem{Fuchs:1999xn}
  J.~Fuchs and C.~Schweigert,
  ``Symmetry breaking boundaries. II: More structures, examples,''
  Nucl.\ Phys.\ B {\bf 568} (2000) 543
  [arXiv:hep-th/9908025].
\bibitem{Fuchs:1999zi}
  J.~Fuchs and C.~Schweigert,
  ``Symmetry breaking boundaries. I: General theory,''
  Nucl.\ Phys.\ B {\bf 558} (1999) 419
  [arXiv:hep-th/9902132].
\bibitem{Quella:2002ns}
  T.~Quella,
  ``On the hierarchy of symmetry breaking D-branes in group manifolds,''
  JHEP {\bf 0212} (2002) 009
  [arXiv:hep-th/0209157].
\bibitem{Klimcik:1996hp}
  C.~Klimcik and P.~Severa,
  ``Open strings and D-branes in WZNW models,''
  Nucl.\ Phys.\ B {\bf 488} (1997) 653
  [arXiv:hep-th/9609112].
\bibitem{Alekseev:1998mc}
  A.~Y.~Alekseev and V.~Schomerus,
  ``D-branes in the WZW model,''
  Phys.\ Rev.\ D {\bf 60} (1999) 061901
  [arXiv:hep-th/9812193].
\bibitem{Gawedzki:1999bq}
  K.~Gawedzki,
  ``Conformal field theory: A case study,''
  arXiv:hep-th/9904145.
\bibitem{Brunner:2005fv}
  I.~Brunner and M.~R.~Gaberdiel,
  ``Matrix factorisations and permutation branes,''
  JHEP {\bf 0507} (2005) 012
  [arXiv:hep-th/0503207].
\bibitem{Enger:2005jk}
  H.~Enger, A.~Recknagel and D.~Roggenkamp,
  ``Permutation branes and linear matrix factorisations,''
  arXiv:hep-th/0508053.
\bibitem{Caviezel:2005th}
  C.~Caviezel, S.~Fredenhagen and M.~R.~Gaberdiel,
  ``The RR charges of A-type Gepner models,''
  arXiv:hep-th/0511078.
\bibitem{DiFrancesco:1997nk}
  P.~Di Francesco, P.~Mathieu and D.~Senechal,
  ``Conformal field theory''. Springer-Verlag, 1997, New York.
\bibitem{Sarkissian:2003yw}
  G.~Sarkissian and M.~Zamaklar,
  ``Symmetry breaking, permutation D-branes on group manifolds: Boundary
  states and geometric description,''
  Nucl.\ Phys.\ B {\bf 696} (2004) 66
  [arXiv:hep-th/0312215].
\bibitem{Elitzur:2001qd}
  S.~Elitzur and G.~Sarkissian,
  ``D-branes on a gauged WZW model,''
  Nucl.\ Phys.\ B {\bf 625} (2002) 166
  [arXiv:hep-th/0108142].
\bibitem{Kubota:2001ai}
  T.~Kubota, J.~Rasmussen, M.~A.~Walton and J.~G.~Zhou,
  ``Maximally symmetric D-branes in gauged WZW models,''
  Phys.\ Lett.\ B {\bf 544} (2002) 192
  [arXiv:hep-th/0112078].
\bibitem{Bardakci:1987ee}
  K.~Bardakci, E.~Rabinovici and B.~Saering,
  ``String Models With $C \leq 1$ Components,''
  Nucl.\ Phys.\ B {\bf 299} (1988) 151.
\bibitem{Gawedzki:1988hq}
  K.~Gawedzki and A.~Kupiainen,
  ``G/H Conformal Field Theory From Gauged WZW Model,''
  Phys.\ Lett.\ B {\bf 215} (1988) 119.
\bibitem{Gawedzki:1988nj}
  K.~Gawedzki and A.~Kupiainen,
  ``Coset Construction From Functional Integrals,''
  Nucl.\ Phys.\ B {\bf 320} (1989) 625.
\bibitem{Karabali:1988au}
  D.~Karabali, Q.~H.~Park, H.~J.~Schnitzer and Z.~Yang,
  ``A Gko Construction Based On A Path Integral Formulation Of Gauged
  Wess-Zumino-Witten Actions,''
  Phys.\ Lett.\ B {\bf 216} (1989) 307.
\bibitem{Karabali:1989dk}
  D.~Karabali and H.~J.~Schnitzer,
  ``Brst Quantization Of The Gauged WZW Action And Coset Conformal Field
  Theories,''
  Nucl.\ Phys.\ B {\bf 329} (1990) 649.
\bibitem{Maldacena:2001ky}
  J.~M.~Maldacena, G.~W.~Moore and N.~Seiberg,
  ``Geometrical interpretation of D-branes in gauged WZW models,''
  JHEP {\bf 0107} (2001) 046
  [arXiv:hep-th/0105038].
\bibitem{Vilenkin:1968nk}
N.~Vilenkin, "Special functions and the theory of group
representations". Providence, R.I., American Mathematical Society
(AMS), (1968) 613p.
















\end{thebibliography}
\end{document}